\documentclass[twocolumn,showpacs,jcp,superscriptaddress]{revtex4}

\usepackage{graphicx}
\usepackage[usenames]{color}
\usepackage{amssymb}
\usepackage{amsmath}
\usepackage{amsfonts}
\usepackage{mathrsfs}

\renewcommand{\epsilon}{\varepsilon}

\begin{document}

\title{Driven polymer translocation through a cylindrical nanochannel: Interplay between the channel length and the chain length}

\author{Huaisong Yong}
\affiliation{CAS Key Laboratory of Soft Matter Chemistry, Department of
Polymer Science and Engineering, University of Science and Technology of
China, Hefei, Anhui Province 230026, P. R. China}

\author{Yilin Wang}
\affiliation{CAS Key Laboratory of Soft Matter Chemistry, Department of
Polymer Science and Engineering, University of Science and Technology of
China, Hefei, Anhui Province 230026, P. R. China}

\author{Shichen Yuan}
\affiliation{CAS Key Laboratory of Soft Matter Chemistry, Department of
Polymer Science and Engineering, University of Science and Technology of
China, Hefei, Anhui Province 230026, P. R. China}

\author{Bi Xu}
\affiliation{CAS Key Laboratory of Soft Matter Chemistry, Department of
Polymer Science and Engineering, University of Science and Technology of
China, Hefei, Anhui Province 230026, P. R. China}

\author{Kaifu Luo}
\altaffiliation{Author to whom the correspondence should be addressed}
\email{kluo@ustc.edu.cn}
\affiliation{CAS Key Laboratory of Soft Matter Chemistry, Department of
Polymer Science and Engineering, University of Science and Technology of
China, Hefei, Anhui Province 230026, P. R. China}

\date{\today}

\begin{abstract}
Using analytical techniques and Langevin dynamics simulations, we investigate the dynamics of polymer translocation through a nanochannel embedded in two dimensions under an applied external field.
We examine the translocation time for various ratio of the channel length $L$ to the polymer length $N$. For short channels $L\ll N$, the translocation time $\tau \sim N^{1+\nu}$ under weak driving force $F$, while $\tau\sim F^{-1}L$ for long channels $L\gg N$, independent of the chain length $N$.
Moreover, we observe a minimum of translocation time as a function of $L/N$ for different driving forces and channel widths. These results are interpreted by the waiting time of a single segment.

\end{abstract}


\maketitle

\section{Introduction}

Polymer translocation through a nanopore or nanochannel is of considerable importance to a multitude of biological functions, such as protein and DNA transport through membrane channels and viral DNA injection into host cells \cite{Alberts}. Recent experiments show that monitoring single molecule translocation through protein or synthetic nanopores has potential applications in single DNA sequencing, single molecular characterization, and single molecular detection \cite{Kasianowicz,Meller03,Akeson,Meller00,Meller01,Meller02,Meller07,Bashir,Sauer,Mathe,Henrickson,Li01,Li03,Li05,
Krasilnikov,Keyser1,Branton,Dekker,Trepagnier,Storm03,Storm05,Storm052}.

The transport of biopolymers through a nanopore has attracted broad interest in the statistical physics community, as it represents a challenging problem in polymer physics \cite{Sung,Muthukumar99,MuthuKumar03,Chuang,Luo1,Luo4,Dubbeldam1,Kantor,Luo2,Matysiak,Dubbeldam2,Panja,Luo3,Aniket,
Slater,Sakaue,Milchev,Guo,LuoMetz,Luo10,Huopaniemi}.
A quantity of particular interest is the average translocation time $\tau$ as a function of the chain length $N$, usually assumed to follow a scaling law $\tau\sim N^{\alpha}$. The scaling exponent $\alpha$ hereby reflects the efficiency of the translocation process.

Standard equilibrium Kramers analysis of diffusion through an entropic barrier yields $\tau \sim N^2$ for the unbiased translocation and $\tau \sim N$ for the driven translocation (assuming
friction is independent of $N$) \cite{Sung,Muthukumar99}.
However, as Chuang \textit{et al}. \cite{Chuang} noted, the quadratic scaling behavior for unbiased translocation cannot be correct for a self-avoiding polymer.
The reason is that the translocation time is shorter than the Rouse equilibration time of a self-avoiding polymer, $\tau_R \sim N^{1+2\nu}$, where the Flory exponent $\nu=0.588$ in three dimensions and $\nu=0.75$ in two dimensions \cite{deGennes}, thus rendering the concept of equilibrium entropy and the ensuing entropic barrier inappropriate for translocation dynamics.
Chuang\textit{et al}. \cite{Chuang} performed numerical simulations with Rouse dynamics for a two dimensional lattice model to study the translocation for both phantom and self-avoiding polymers.
Their results show that for large $N$ translocation time $\tau \sim N^{1+2\nu}$, which scales approximately in the same manner as the equilibration time but with a much larger prefactor. For driven translocation, Kantor and Kardar \cite{Kantor} have demonstrated that the assumption of equilibrium in polymer dynamics by Sung and Park \cite{Sung}
and Muthukumar \cite{Muthukumar99} breaks down more easily and provided a lower bound $\tau \sim N^{1+\nu}$ for the translocation time by comparison to the unimpeded motion of the polymer. Most recently, we have found that for faster translocation processes $\alpha=1.37$ in three dimensions \cite{Luo7}, while it crosses over to $\alpha=1+\nu$ for slower translocation, corresponding to weak driving forces and/or high friction \cite{Luo7}.

Translocation process can be affected by many factors, such as the driving force, and the shape and the size of the pore. As to the driving force, it can be provided by an electric field, a chemical potential, a pulling force, or geometrical confinement of the polymer.

However, most of previous studies focus on short nanopores, where the pore length is much smaller than the radius of gyration of the polymer. The dynamics of polymer translocation through a long nanochannel
is of great importance, because it is related to many technological applications, such as the ultrafiltration process \cite{Chi}. Although polymer translocation through a cylindrical channel of finite diameter and length between two spherical compartments has been theoretically investigated \cite{Kolomeisky,Muthukumar08}, equilibrium entropy of polymer as a function of the position during the translocation is used, leading to the results to be correct at most for Gaussian chain. The translocation dynamics for polymer through a long channel is still not clear due to the interplay of the chain length and the length of the channel. To this end, using both analytical techniques and Langevin dynamics (LD) simulations, we investigate the dynamics of polymer translocation through a long nanochannel.

The paper is organized as follows: in Section II, we describe our simulation model and methods; in Section III, we present the results of blob and scaling theories with discussing our simulation data of polymer chain translocation process; the conclusions are drawn in Section IV.

\section{Model and methods} \label{chap-model}

\begin{figure}
\includegraphics[width=8.2cm]{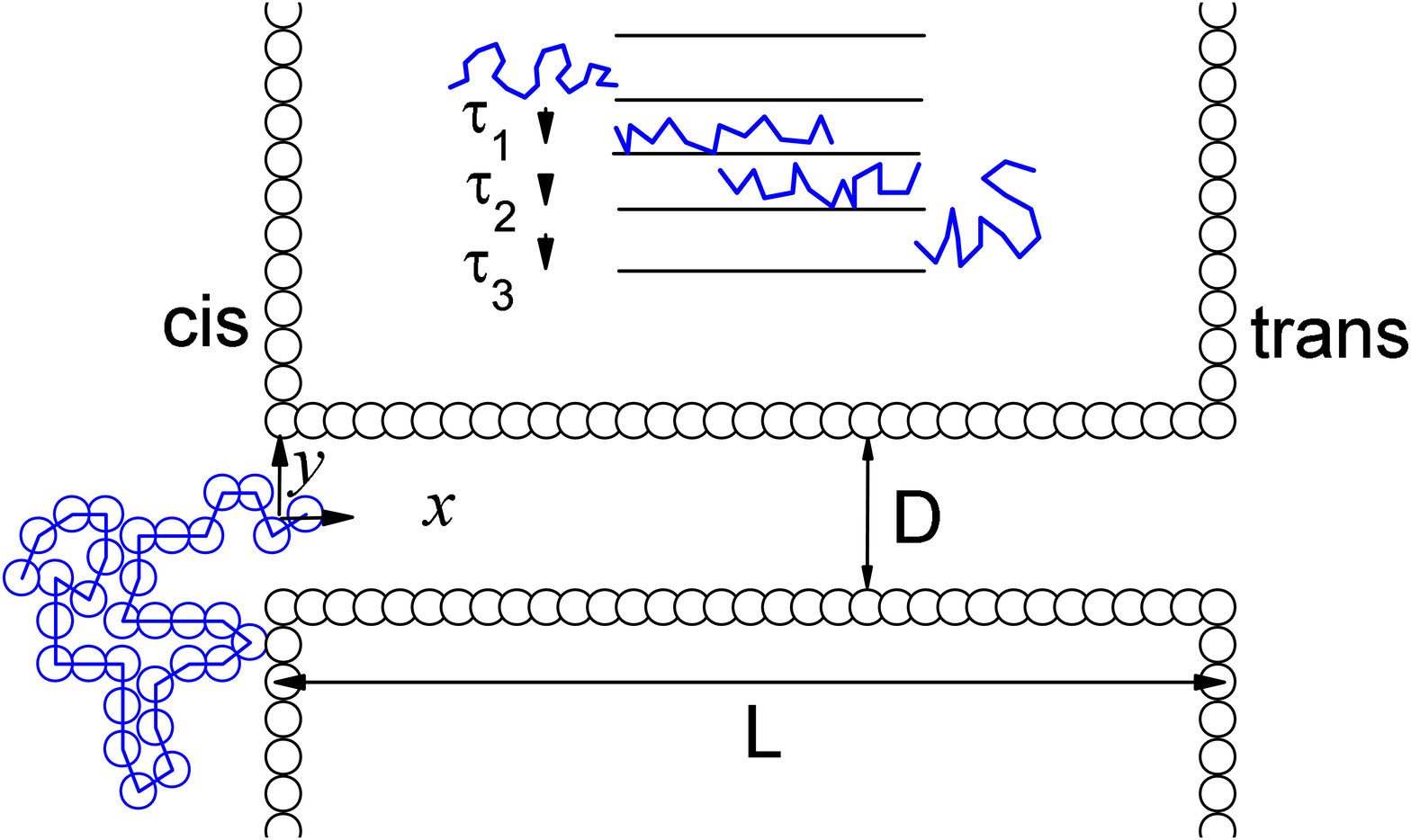}
\caption{Schematic representation of polymer translocation through a two-dimensional long nanochannel of length $L\sigma$ and width $D\sigma$. The driving force acting on the bead in the channel can be provided by an applied intra-channel electric field.
For long nanochannels, the translocation process can be broken into three components $\tau\approx \tau_{1}+\tau_{2}+\tau_{3}$, where $\tau_{1}$, $\tau_{2}$ and $\tau_{3}$ correspond to initial filling of the nanochannel, transfer of the polymer from the starting \textit{cis} side to the \textit{trans} side, and finally emptying of the nanochannel, respectively. }
\label{Fig1}
\end{figure}

We consider the two dimensional geometry where the nanochannel is sandwiched between two parallel walls that are close to each other with the polymer chain, as shown in Fig. \ref{Fig1}.
In our simulations, the polymer chains are modeled as bead-spring chains of Lennard-Jones (LJ) particles with the Finite Extension Nonlinear Elastic (FENE) potential \cite{Kremer}. Excluded volume interactions between beads are taken into account by a short range repulsive LJ potential
\begin{equation}
U_{\mathrm{LJ}} (r)=\left\{  \begin{array}{ll}  4\epsilon \left[ \left(\sigma/r\right)^{12}-\left(\sigma/r\right)^6  \right]+\epsilon, & r\le 2^{1/6}\sigma\\
0, &  r>2^{1/6}\sigma,
\end{array}\right.
\end{equation}
where $\sigma$ is the diameter of a bead and $\epsilon$ is the potential depth. The connectivity between neighboring beads is modeled as a FENE spring with

\begin{equation}
U_{\mathrm{FENE}}(r)=-\frac{1}{2}kR_0^2\ln\left(1-r^2/R_0^2\right),
\end{equation}
where $r$ is the distance between consecutive beads, $k$ is the spring constant, and $R_0$ is the maximum allowed separation between connected beads.

As shown in Fig. \ref{Fig1}, the wall is formed by stationary particles with the diameter $\sigma$. In unit of $\sigma$, the nanochannel has length $L$ and width $D$.
Between all bead-wall particle pairs, there exists the same short range repulsive LJ interaction as described above. In the Langevin dynamics (LD) simulations, each bead is subjected to conservative, frictional, and random forces, respectively, with \cite{Allen}

\begin{equation}
m\mathbf{\ddot{r}}_i=-\mathbf{\nabla}({U}_{\mathrm{LJ}}+{U}_{\mathrm{FENE}})+
\mathbf{F}_{\mathrm{ext}}-\xi\mathbf{v}_i+\mathbf{F}_i^R,
\end{equation}
where $m$ is the bead mass, $\xi$ is the friction coefficient for a single bead, $\mathbf{v}_i=\mathbf{\dot{r}}_i$ is the bead velocity, and $\mathbf{F}_i^R$ is the random force satisfying the fluctuation-dissipation theorem. The external force is expressed as $\mathbf{F}_{\mathrm{ext}}=F\hat{x}$, where $F$ is the strength of force exerted on the translocating beads located inside the channel, and $\hat{x}$ is a unit vector in the direction along the channel. Experimentally, this external driving force acting on the bead in the channel can be provided by an applied transmembrane electric field.

In the present work, we use the LJ parameters $\epsilon$ and $\sigma$ and the bead mass $m$ to fix the energy, length and mass scales, respectively. The time scale is then given by $t_{\mathrm{LJ}}=(m\sigma^2/\epsilon)^{1/2}$. The dimensionless parameters in our simulations are $\xi=0.7$ and $k_{B}T=1.2$ unless otherwise stated. For dimensionless parameters for FENE potential, we use $R_0=2$ and $k=7$ \cite{Limbach}. We have also checked that using $R_0=1.5$ and $k=30$ does not change our results.
The Langevin equation is integrated in time by a method described by Ermak and Buckholz \cite{Ermak} in two dimensions.

Initially, the first bead of the chain is placed just inside the channel (at $x=0.75$, $y=0$), while the remaining beads are under thermal collisions described by the Langevin thermostat to obtain an equilibrium configuration. The translocation time is defined as the time interval between the entrance of the first segment into the pore and the exit of the last segment. Typically, we average the data over 1000 independent runs.

\section{Result and Discussion} \label{chap-results}

\subsection{Theory}

For translocation through short channels, the translocation time $\tau$ has been estimated by Kantor \textit{et al.} \cite{Kantor} using the scaling arguments, where an essential assumption is that the chain is not severely deformed during translocation and in particular the chain configurations at both \textit{cis} and \textit{trans} sides are close to equilibrium. For slow translocation, these assumptions are satisfied. Thus, the translocation time can be written as \cite{Kantor}
\begin{equation}
\tau\sim R_g/v \sim N^{1+\nu}/F
\label{eq4}
\end{equation}
for $L\ll R_g$.
Here, $L$ is the channel length, $R_g\sim N^\nu$ is the radius of gyration of the polymer without the confinement, and $v=F/(N\xi)$ is the average translocation velocity. The predicted exponent $1+\nu$ is observed in the slow dynamics regime \cite{Luo7}.

For translocation through long channels, we can break down the translocation process into three components, as shown in Fig. \ref{Fig1}. The total translocation time can be written as a sum of three contributions $\tau\approx \tau_{1}+\tau_{2}+\tau_{3}$, where $\tau_{1}$, $\tau_{2}$ and $\tau_{3}$ correspond to initial filling of the nanochannel, transfer of the polymer from the starting \textit{cis} side to the \textit{trans} side, and final emptying of the nanochannel, respectively.
Particularly, for an infinite long channel of length $L$, the translocation dynamics is dominated by $\tau_2$, which can be written as $\tau_2 \sim L/v$, with $v$ being the average translocation velocity. Based on the balance of the frictional force and the driving force, $N\xi v=NF$ where $F$ is the driving force acting on one bead, we have $v=F/\xi$. Therefore, the translocation time
\begin{equation}
\tau \approx \tau_2 \sim L/v \sim L\xi /F
\label{eq5}
\end{equation}
for $L\gg R_g$. This result indicates that $\tau$ is independent of the chain length $N$.

Based on Eqs. (\ref{eq4}) and (\ref{eq5}), there exists a crossover for the scaling behavior of $\tau$ as a function of $L$. For short channels $L\ll N$, the translocation time $\tau \sim N^{1+\nu}$ under weak driving force $F$ \cite{Luo7}, while $\tau\sim F^{-1}L$ for long channels $L\gg L$.
However, $\tau$ is still not clear for various scaled pore length $L/N$. To this end, in the following we give more details from simulations.

\subsection{Numerical results}

In our simulations, we find that the translocation probability is approximately independent of the chain length $N$ as observed in the previous study \cite{LuoMetz}. This result is also in agreement with the experiments \cite{Cabodi,Mannion}, where the polymer chain insertion process is entirely local to the portion of the chain at the interface and independent of the chain length.

\begin{figure}
\includegraphics[width=8.2cm]{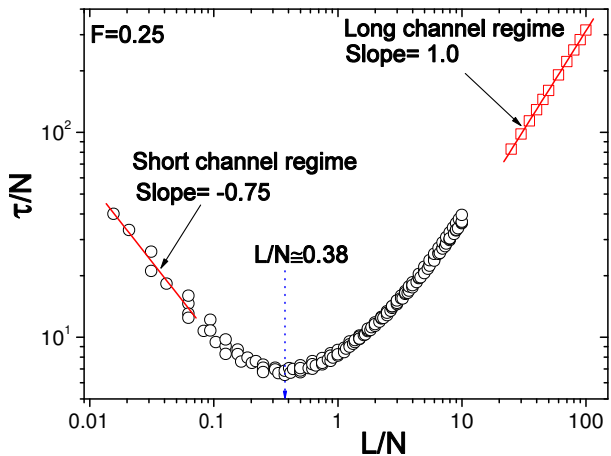}
\caption{$\tau/N$ as a function of $L/N$ for the channel width $D=3$ and the driving force $F=0.25$. Here, we use $N=32$, 64, 96 and 128.}
\label{Fig2}
\end{figure}

\begin{figure}
\includegraphics[width=8.2cm]{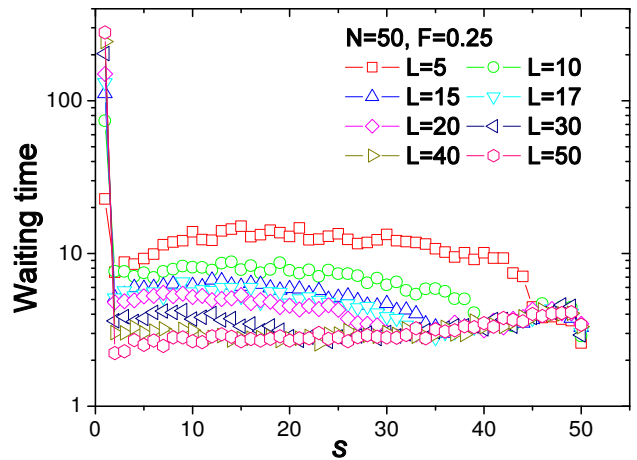}
\caption{Waiting time distribution for $N=50$, $D=3$, $F=0.25$ and different channel length $L$. Here, $t_{s=1}$ includes the the time duration for initial filling process $\tau_1$.}
\label{Fig3}
\end{figure}

\begin{figure}
\includegraphics[width=8.2cm]{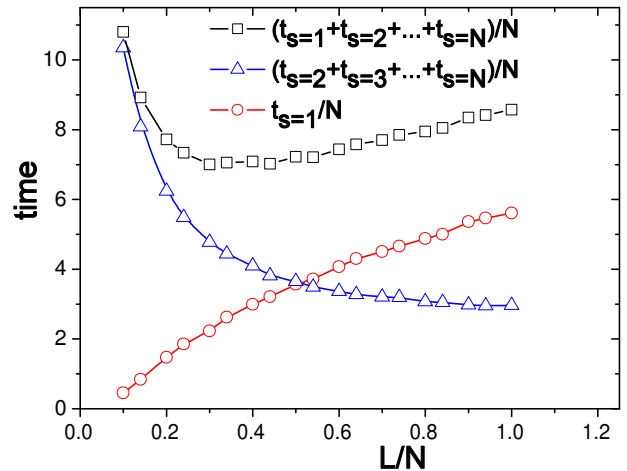}
\caption{Sum of waiting times for all beads ($\sum_{i=1}^{N}t_{s=i}/N$), sum of waiting times for beads numbered from 2 to $N$ ($\sum_{i=2}^{N}t_{s=i}/N$) and waiting time for the first bead $t_{s=1}/N$ as a function of $L/N$ for $N=50$, $D=3$ and $F=0.25$. Here, $t_{s=1}$ includes the the time duration for initial filling process $\tau_1$.}
\label{Fig4}
\end{figure}

A particularly interesting question concerns $\tau$ for various $L/N$. Fig. \ref{Fig2} shows the scaled translocation time $\tau/N$ as a function of $L/N$ for $F=0.25$. For $L/N \le 0.1$, we find $\tau/N \sim (L/N)^{-\nu}$ with $\nu=0.75$ being the Flory exponent in two dimensions. This implies $\tau \sim N^{\alpha}$ with $\alpha={1+\nu}$, in agreement with the prediction in Eq. (\ref{eq4}) for $L\ll R_g$.

In our previous results, we find that with increasing the driving force $F$ the exponent $\alpha$ crosses over from ${1+\nu}$ to $2\nu$ in two dimensions \cite{Luo2} and from ${1+\nu}$ to 1.37 in three dimensions \cite{Luo7}. Due to translocation through a quite short pore with a small $L$, the driving force acting on the chain is quite weak, and thus the exponent ${1+\nu}$ is in agreement with the previous result for slow dynamics regime.
However, in contrast to our results \cite{LuoMetz,Luo2}, recently Lehtola \textit{et al.} \cite{Lehtola} and Dubbeldam \textit{et al.} \cite{DubbeldamEprint} also observe the crossover exponent from $2\nu$ to ${1+\nu}$ with increasing $F$. This discrepancy may be from the chain length and the driving force used in the simulations \cite{Ikonen}.

When $L/N \ge 10$, we observed $\tau/N\sim L/N$, which demonstrates that $\tau \sim L$, independent of the chain length. This scaling behavior is in agreement with the prediction in Eq. (\ref{eq5}) for $L\gg R_g$.
Therefore, the observed crossover scaling behavior is in good agreement with our theoretical predictions in Eqs.  (\ref{eq4}) and (\ref{eq5}).

The most interesting result in Fig. \ref{Fig2} is that the translocation time has a minimum for an ¡°optimal¡± value of $L/N$ at $L/N\approx 0.38$.
To understand the existence of the minimum, we take into account the dynamics of a single segment passing through the channel during translocation. We numerically calculated the waiting times for all beads in a chain. The waiting time of bead $s$ for successful translocation
hereby is defined as the average time between the events that bead $s$ and bead
$s+1$ exit the channel. Fig. \ref{Fig3} shows the waiting time distribution for $N=50$, $D=3$, $F=0.25$ and different channel length $L$. For $L>5$, it takes much longer time for the first segment to exit the channel, denoted as $t_{s=1}$. \textit{Here, we should point out that $t_{s=1}$ includes the time duration for initial filling process $\tau_1$ because the first segment must pass through the whole channel before its exiting the channel}.
Moreover, $t_{s=1}$ increases with increasing $L$, while the waiting times for other monomers ($t_{s=2}$, $t_{s=3}$...$t_{s=N}$) initially decrease and then saturate with increasing $L$. Quantitatively, we measure the sum of waiting times for all beads $\sum_{i=1}^{N}t_{s=i}/N$, sum of waiting times for beads numbered from 2 to $N$, $\sum_{i=2}^{N}t_{s=i}/N$, and the waiting time for the first bead $t_{s=1}/N$ as a function of $L/N$, see Fig. \ref{Fig4}. With increasing $L/N$, $t_{s=1}/N$ first increases rapidly followed by a slower increases, while $\sum_{i=2}^{N}t_{s=i}/N$ initially decreases rapidly and then almost saturates for long channels. The interplay between $t_{s=1}/N$ and $\sum_{i=2}^{N}t_{s=i}/N$ leads to the observed minimum of the translocation time (the curve for $\sum_{i=1}^{N}t_{s=i}/N$). Recently, this effect was also observed in unbiased translocation through long channel based on a computer simulation \cite{Luo1}, and was predicted for translocation through a long channel between two spherical compartments \cite{Muthukumar08}.

To generalize above observed results, we examine $\tau/N$ as a function of $L/N$ for different driving forces and channel widths in the following.

\subsubsection{The influence of the driving force on the translocation time}

\begin{figure}
\includegraphics[width=8.2cm]{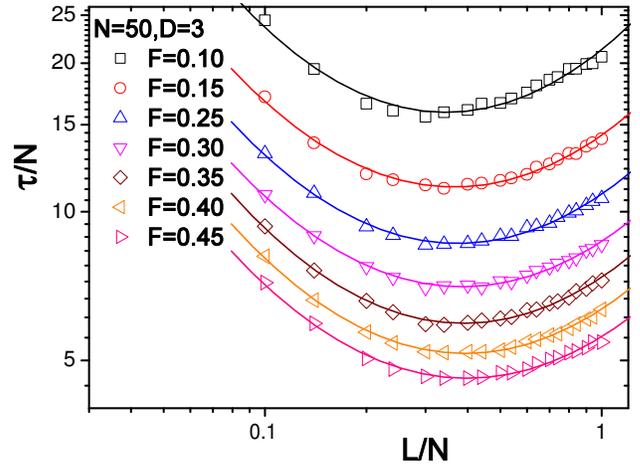}
\caption{$\tau/N$ as a function of $L/N$ for $N=50$, $D=3$ and different driving forces. }
\label{Fig5}
\end{figure}

\begin{figure}
\includegraphics[width=8.2cm]{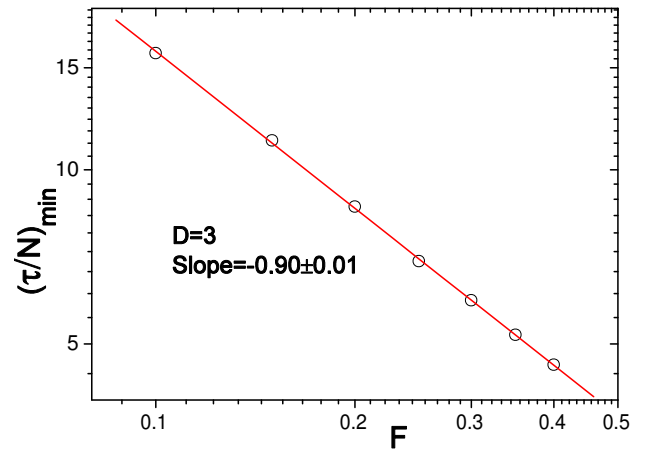}
\caption{The minimum of the translocation time as a function of the driving force for $N=50$ and $D=3$.}
\label{Fig6}
\end{figure}

Fig. \ref{Fig5} shows  $\tau/N$ as a function of $L/N$ for $N=50$, $D=3$ and different driving forces. As expected, with increasing $F$, $\tau/N$ decreases for the range of $L/N$ we examined. Moreover, we observe the minimum of the translocation time, $(\tau/N)_{min}$, for different $F$. The critical ratio, $(L/N\sigma)_c$, corresponding to the minimum of $\tau/N$, almost does not change with $F$.

\begin{figure}
\includegraphics[width=8.2cm]{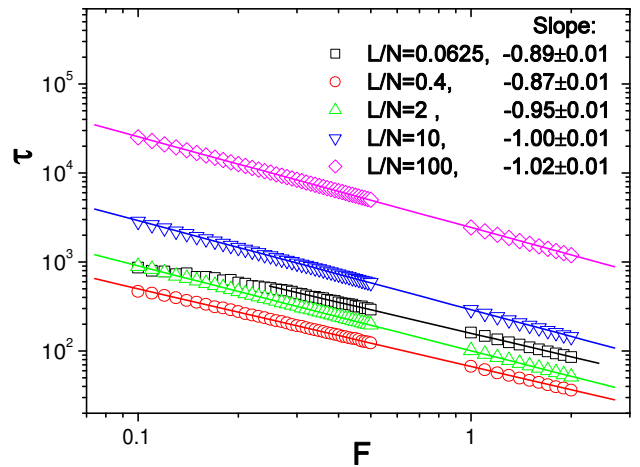}
\caption{Total translocation time $\tau$ as a function of  the driving force $F$ for different $L/N$. The chain length $N=32$.}
\label{Fig7}
\end{figure}

As shown in Fig. \ref{Fig6}, we further find $(\tau/N)_{min}\sim F^{\beta}$, with $\beta={-0.90\pm0.01}$. This exponent is close to -1.
Fig. \ref{Fig7} shows the translocation time $\tau$ as a function of the driving force $F$ for different $L/N$. $\tau$ initially decreases and then increases with increasing $L/N$, which further confirms that $\tau$ has minima at $L/N \approx 0.4$ for different driving forces $F$. In addition, we also observe $\tau\sim F^{-1}$ for long channels with $L/N \ge 2.0$. For $L/N \le 0.4$, exponents are a little larger than -1. Particularly, for $L/N = 0.0625$ and $F\le 0.2$, the driving force is very weak and the translocation is almost controlled by pure diffusion as demonstrated in Ref. \onlinecite{Huopaniemi}, leading to a weaker dependence of $\tau$ on $F$.

\subsubsection{The influence of the channel width on the translocation time}

\begin{figure}
\includegraphics[width=8.2cm]{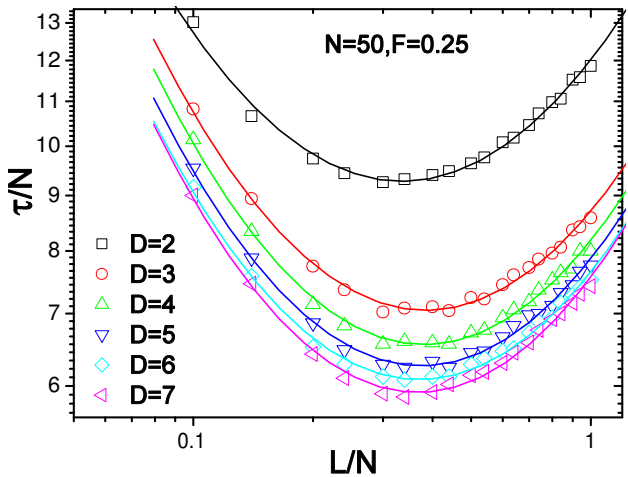}
\caption{$\tau/N$ as a function of $L/N$ under the driving force $F=0.25$ for the chain length $N=50$ and different channel widths $D$.}
\label{Fig8}
\end{figure}

\begin{figure}
\includegraphics[width=8.2cm]{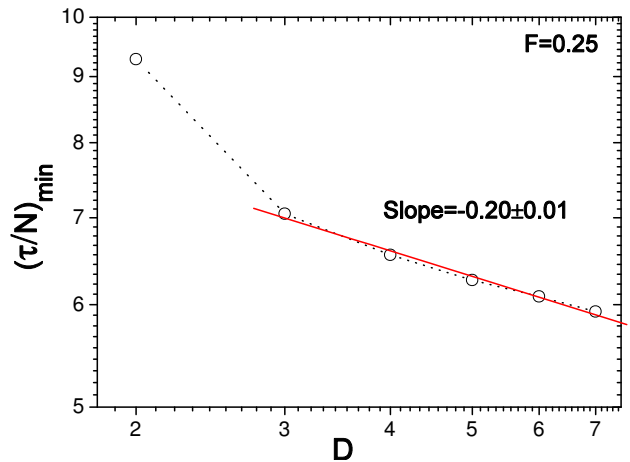}
\caption{The minimum of the translocation time as a function of the channel width $D$ under the driving force $F=0.25$ for $N=50$.}
\label{Fig9}
\end{figure}

\begin{figure}
\includegraphics[width=8.2cm]{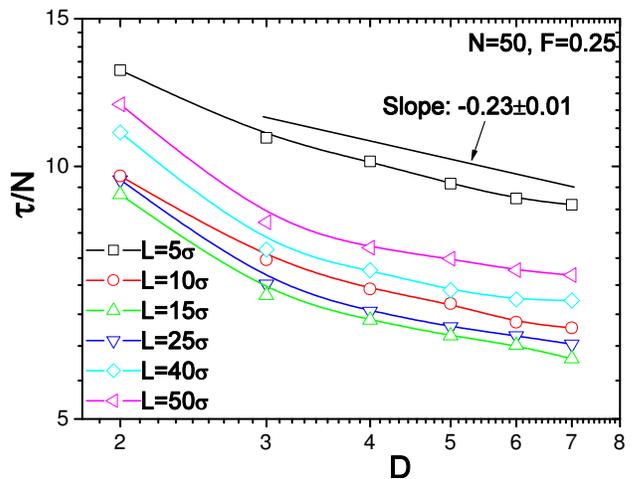}
\caption{$\tau/N$ as a function of the channel width $D$ for $N=50$ and $F=0.25$.}
\label{Fig10}
\end{figure}

$\tau/N$ as a function of $L/N$ under the driving force $F=0.25$ for $N=50$ and different $D$ is shown in Fig. \ref{Fig8}, where $\tau$ also shows the minimum.
The critical ratio, $(L/N)_c$, corresponding to the minimum of $\tau/N$, only has a very weak $D$ dependence based on the fitted curves. This behavior is a little different from the results for non-driven translocation through a long channel between two spherical compartments \cite{Muthukumar08}, where the critical channel length corresponding to the minimum of $\tau$ always increases with increasing the channel width. This is because the entropic force for polymer partially inside the channel depends on $D$ for unbiased case.

In addition, $\tau/N$ decreases with increasing $D$ for different $L/N$. This is because the channel can accommodate more beads with increasing $D$, leading to larger driving force.
The minimum of the translocation time, $(\tau/N)_{min}$ decrease rapidly first, and then is slowed down with increasing $D$, see Fig. \ref{Fig9}. For $D\ge 3$, we observe $(\tau/N)_{min}\sim D^{-0.20\pm0.01}$. Fig. \ref{Fig10} shows the $\tau/N$ as a function of the channel width $D$ for $N=50$ and $F=0.25$.
$\tau/N$ has a weaker dependence on $D$ compared with the equilibrium longitudinal size of the chain confined to a channel, $ R_{\parallel}\sim N\sigma(\frac{\sigma}{D})^{1/\nu_{2}-1}\sim ND^{-1/3}$ \cite{deGennes}. This may be due to the elongations of the chain inside the channel under the driving force and the non-equilibrium process of the translocation.

\section{Conclusions}
\label{chap-conclusions}

Using analytical techniques and Langevin dynamics simulations, we investigate the dynamics of polymer translocation through a nanochannel embedded in two dimensions under an applied external field.
We examine the translocation time for various ratio of the channel length $L$ to the contour length of polymer $N$. For short channels $L\ll N$, the translocation time $\tau \sim N^{1+\nu}$ under weak driving force $F$, while $\tau\sim F^{-1}L$ for long channels $L\gg N$, independent of the chain length $N$.
Moreover, we observe a minimum of translocation time as a function of $L/N$ for different driving forces and channel widths. These results are interpreted by the waiting time of a single segment.
This work is relevant for the applied biophysics field. Our results can help design a device geometry to achieve higher separation resolution for DNA molecules using nanochannel.

In our previous study \cite{Luo10}, we have examined the different case where the driving force acts solely on the bead at the entrance of the channel. Due to the crowding effect induced by the partially translocated monomers, the translocation dynamics is significantly altered in comparison to an unconfined environment. The scaling exponent of the translocation time $\tau$ with the chain length $N$ depends on the channel width and the driving force.

\begin{acknowledgments}
This work is supported by the National Natural Science Foundation of China (Grant Nos. 21074126, 21174140, J1030412), and the ``Hundred Talents Program'' of Chinese Academy of Science (CAS).
\end{acknowledgments}

\includegraphics*[width=8cm]{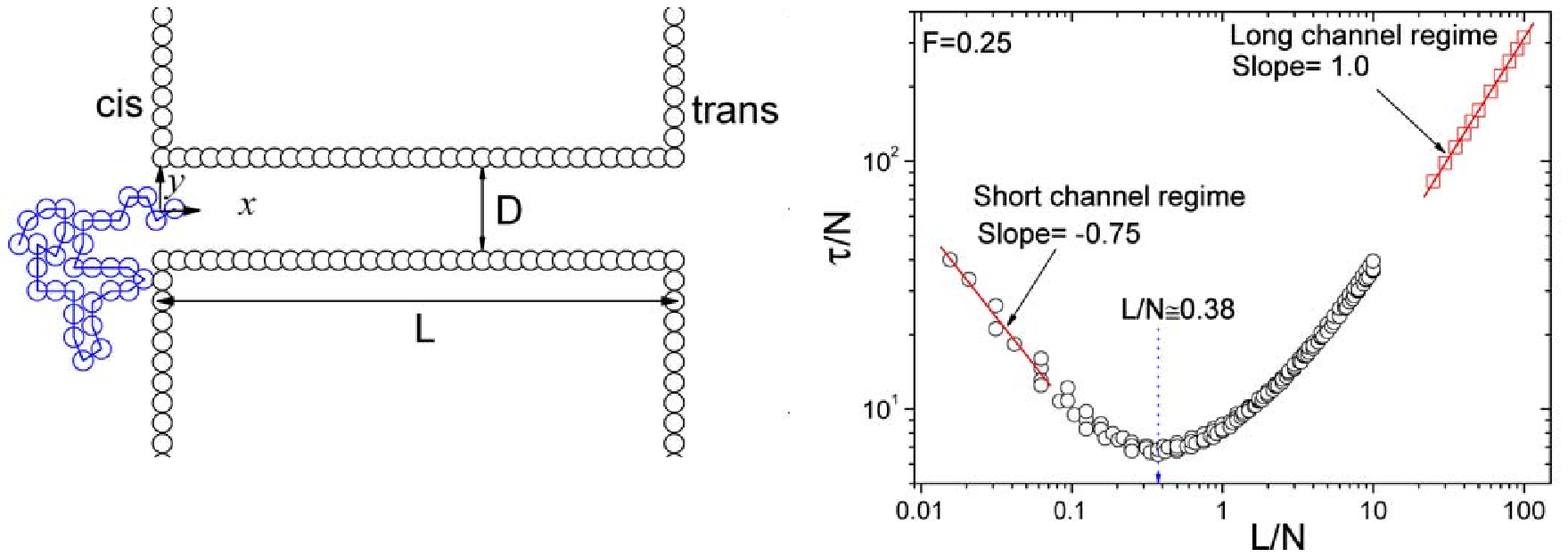}
``Translocation time as a function of the channel length shows a minimum for different driving forces and channel widths."

\end{document}